\documentclass[12pt,reqno]{article}

\usepackage{amsfonts}
\usepackage{amsmath}
\usepackage{amssymb}
\usepackage{amsthm}
\usepackage{setspace}
\usepackage{fullpage}
\usepackage{graphicx}
\usepackage{url}
\usepackage{braket}
\usepackage{mathtools}
\usepackage[all]{xy}
\usepackage{caption}

\usepackage{color}
\usepackage{bm}

\usepackage[numbers]{natbib}

\usepackage{verbatim}
\usepackage{graphicx}

\usepackage{mathrsfs}

\theoremstyle{remark}

\theoremstyle{definition}

\newcommand{\ZZ}{\mathbb{Z}}
\newcommand{\HH}{\mathbb{H}}
\newcommand{\RR}{\mathbb{R}}
\newcommand{\CC}{\mathbb{C}}

\newcommand{\del}{\partial}
\newcommand{\delbar}{\bar{\partial}}

\newcommand{\tensor}{\otimes}

\newcommand{\oo}{\infty}
\DeclareMathOperator{\tr}{tr}

\DeclareMathOperator{\vol}{vol}

\newcommand{\PP}{\mathbb{P}}

\newcommand{\wt}{\widetilde}

\newcommand{\Spin}{\text{Spin}}
\newcommand{\String}{\text{String}}
\newcommand{\Pin}{\text{Pin}}
\newcommand{\pt}{\text{pt}}

\begin{document}

\begin{titlepage}
\hfill \\
\vspace*{15mm}
\begin{center}
{\Large \bf Cobordism Classes and the Swampland}

\vspace*{15mm}

{\large Jacob McNamara and Cumrun Vafa}
\vspace*{8mm}

Jefferson Physical Laboratory, Harvard University, Cambridge, MA 02138, USA\\

\vspace*{0.7cm}
%%\maketitle

\end{center}
\begin{abstract}

We argue that any proposed quantum theory of gravity with non-trivial cobordism classes in the space of configurations belongs to the Swampland. The argument is based on the assumption that there are no global symmetries in a consistent theory of quantum gravity. The triviality of the cobordism classes requires the existence of certain stringy defects that trivialize the potential cobordism classes. We provide evidence for this conjecture by identifying those defects demanded by this argument that could preserve supersymmetry, and predict the existence of new non-supersymmetric defects in string theory.

\end{abstract}

\end{titlepage}

\tableofcontents

\section{Introduction}

One of the main lessons of the second superstring revolution is that all known string theories are connected by a web of dualities. As a result of this discovery, we have come to think of string theory as a single theory of quantum gravity rather than many different, independent theories. It is important to note, however, that many of these dualities involve compactification. For example, while it doesn't appear that Type IIA and Type IIB are equivalent theories in 10 dimensions, T-duality relates their compactifications on $S^1$ from 10 to 9 dimensions. As another example, IIA compactified to 6 dimensions on $K3$ is equivalent to the heterotic theories compactified on $T^4$, even though they are not currently expected to be related in 10 dimensions.

In what sense, then, does this mean that string theory is a unique theory of quantum gravity? The precise thing to say is that string dualities suggest the uniqueness of string theory defined on fully compact space, leaving only time noncompact. In fact, there is another sense in which string theory is not unique in infinite space, which is the presence of a landscape of vacua for compactifications to $d \geq 3$ dimensions. Because it would require infinite energy to change the moduli throughout all of space, different choices of moduli define superselection sectors, which could equally well be thought of as different theories of quantum gravity. However, this is not the language normally used, and for a good reason, since localized regions can fluctuate between different moduli at a finite energy cost. Thus, it is not the bulk theory which is non-unique, but merely the boundary conditions at infinity.\footnote{ Even if one fixes the amount of supersymmetry, it has been suggested that in some cases each different choice of the topology is connected by allowed processes in string theory, such as the conjecture of Reid \cite{Reid} that all Calabi-Yau threefolds are connected by geometric transitions.}

Motivated by the example of a moduli space of vacua, we can ask whether, say, IIA and IIB in 10 dimensions are connected in the above sense, in that there could exist a finite energy domain wall between the two. This is a stronger notion of connectedness than that given by T-duality, since though the 10 dimensional Type II theories are connected in the moduli space of their compactifications on $S^1$, they are at infinite distance from each other.  As we will show in Section \ref{New Predictions}, any domain wall between IIA and IIB would necessarily break all supersymmetry. Since many of the ideas we have developed in string theory rely on supersymmetry, it is possible that we would have missed the existence of such non-supersymmetric configurations, whose construction may require harnessing non-perturbative corrections in non-supersymmetric situations.

To analyze this stronger notion of connectedness, we are led to consider the cobordism of quantum gravity. What we mean by this is the following. Consider the collection of all configurations of a theory of quantum gravity with $D$ large spacetime dimensions. We can define an equivalence relation on this collection by declaring two configurations to be equivalent if they are connected as above by finite energy domain walls. We will write $\Omega^{\text{QG}, D}$ for the set of equivalence classes. In fact, $\Omega^{\text{QG}, D}$ carries an abelian group structure, coming from stacking theories, as we will discuss in Section \ref{Definition of Cobordism}, and so we will call this group the cobordism group of quantum gravity with $D$ large dimensions. 

In order to justify the term ``cobordism" for this group, suppose our theory of quantum gravity with $D$ large dimensions arises as the compactification of a theory in $d$ total dimensions, compactified on a space of dimension $k = d - D$ (as in the case of string compactifications). Then we can replace $\Omega^{\text{QG}, D} $ with $ \Omega_k^\text{QG}$,
where $\Omega_k^\text{QG}$ denotes the cobordism groups of $k$-dimensional compactification spaces, since a nontrivial cobordism between two compactification spaces behaves as a domain wall between the two $D$-dimensional theories, and vice versa. If quantum gravity only made sense on smooth manifolds, then this would agree with the usual mathematical notion of cobordism of smooth manifolds with some specified geometric structure. While quantum gravity makes sense on spaces much more general than smooth manifolds, including singular spaces and non-geometric backgrounds, the notion of cobordism of quantum gravitational backgrounds should still make sense. Indeed, the more physically well motivated cobordism groups $\Omega^{\text{QG}, D}$ make no reference to the geometry (or lack thereof) of the internal dimensions. However, for the rest of this paper, we will use the notation $\Omega_k^\text{QG}$ as opposed to $\Omega^{\text{QG}, D}$, since we will be comparing to results from the study of cobordisms of the manifolds used in string compactifications.

In this paper, we argue that the presence of a nontrivial cobordism group $\Omega_k^\text{QG}$ signals an inconsistency of the $d$-dimensional theory, and thus places it in the Swampland. In particular, we will argue that, were a theory of quantum gravity to have a nontrivial cobordism group $\Omega_k^\text{QG} \neq 0$, then there would be a global $(d - k - 1)$-form symmetry \cite{Gaiotto:2014kfa}, with charges labeled by classes
\[ [M] \in \Omega_k^\text{QG}. \]
Since we believe that all theories of quantum gravity arising from compactifications of string theory are free from any global symmetry, this implies that any theory in the string landscape has $\Omega_k^\text{QG} = 0$. In fact, in the context of AdS/CFT, the absence of global symmetries in the bulk can be established on more rigorous grounds \cite{Harlow:2018tng}, and so in that context the condition $\Omega_k^\text{QG} = 0$ should follow as well.

There is one caveat to this argument, which is that in sufficiently low dimension, global symmetries do seem to be allowed in quantum gravity. One class of examples are the string worldsheets, viewed as 2d quantum gravitational theories, where global symmetries are plentiful. In fact, in these cases, there exist nontrivial cobordism invariants \cite{Kaidi:2019pzj}, which can be used to differentiate distinct types of string worldsheets, and thus show the non-uniqueness of 2d quantum gravity theories.  If we were to be conservative (and avoid the special properties of low dimensional quantum gravity), then we would restrict ourselves to claiming only that $\Omega_k^\text{QG}$ should vanish for $k \geq 3$, and indeed this is the restriction for the black hole argument in Section \ref{Conserved Charges and Black Holes} to apply. However, it is not clear to what extent we should consider low dimensional theories like the string worldsheet as part of the landscape of target space string theory, and so we will assume for the rest of this paper that all global symmetries are indeed forbidden in the string landscape.

The condition $\Omega_k^\text{QG} = 0$ implies the existence of a number of objects in string theory, some known and some currently unknown. In particular, it implies that for any valid compactification of string theory (or any quantum gravity theory) on a $k$-dimensional space, there is a compactification on a $(k+1)$-dimensional space that has our original compactification as its boundary. Viewing the bulk of this $(k+1)$-dimensional compactification as a $(d - k - 1)$-dimensional defect, we see that the vanishing of cobordism groups implies the existence of certain defects in string theory.  
These defects do not have to be independent dynamical objects; for example, they could involve junctions of dynamical branes, or smooth manifolds on which some other structure degenerates.   As we will see, these defects include many of the known objects in string theory, including D-branes and O-planes. However, for some examples, the required defects are not currently known in string theory. In all of those examples where we cannot identify the required defect, we can show that they must break all supersymmetry, and so using topological arguments we are able to predict that consistency of quantum gravity requires certain non-supersymmetric defects to exist in string theory!\footnote{ The alternative, that string theory does in fact have some global symmetries, is theoretically possible but is in tension with a great deal of evidence from many sources. } In a sense this is a quantum gravitational analog of the recent insights into the dynamics of non-supersymmetric QFT's gained using topological arguments.

From the perspective of the $D$-dimensional macroscopic spacetime, the condition of triviality of cobordism groups is equivalent to saying that it should be possible to create, with finite energy, a void in the space, where the $(D-1)$-dimensional space ends on a $(D-2)$-dimensional boundary. In particular we predict that in our universe we should be able to create finite-energy bubbles whose interior are forbidden regions (i.e., do not exist)!  Of course, by putting domain walls back to back, this also implies that we can obtain any other allowed gravitational theory inside the bubble as well.\footnote{Whether these bubbles would be hidden behind event horizons depends on the size of the bubble and the tension of the domain wall.}

This paper is organized as follows. In Section \ref{Definition of Cobordism}, we define the cobordism groups $\Omega_k^\text{QG}$. In Section \ref{Cobordism Invariants in Quantum Gravity}, we argue from black hole physics and the absence of global symmetries in quantum gravity that $\Omega_k^\text{QG} = 0$, and discuss how this condition can be used to predict structure in a theory of quantum gravity. In Section \ref{Examples}, we examine the vanishing of cobordism groups in some restricted approximations to string theory. In Section \ref{New Predictions}, we discuss some of the new non-supersymmetric defects in string theory, predicted by the condition of vanishing cobordism groups. Finally, in Section \ref{Conclusion} we conclude with some ideas for future work.

\section{Definition of Cobordism}\label{Definition of Cobordism}

Cobordism can refer to a number of increasingly refined mathematical objects, but at its most basic form, cobordism is an equivalence relation between smooth manifolds more general than diffeomorphism. While two diffeomorphic manifolds are equivalent on the point-set level, two manifolds that are cobordant are related by a sequence of allowed topology-changing operations, and so is the natural notion of equivalence in order to describe the connectedness of quantum gravity as discussed in the introduction. In this section, we define what we mean by the cobordism groups $\Omega_k^\text{QG}$ of quantum gravity.

\subsection{Cobordism Groups of Quantum Gravity}

Suppose we have a quantum gravity theory defined in $d$ spacetime dimensions, and we are considering compactifications of the theory to $D = d - k$ dimensions. If we are in the situation where the internal dimensions are geometric, this compactification can be described by some compact $k$-dimensional manifold $M^k$ on which the theory is compactified, and our notion of cobordism reduces to the usual mathematical notion of cobordism. Since we may have more general notions of $k$-dimensional backgrounds of quantum gravity, we need a more general definition.

As discussed above, we would like to say that two $k$-dimensional backgrounds $M, N$ of quantum gravity are cobordant, $M \sim N$, if they are related by a sequence of allowed topology-changing processes. If we place no restrictions at all on what topology changes are allowed, than all backgrounds are trivially related to all other backgrounds, simply because we have been too general about what we mean by ``topology change." The relevant notion of ``allowed topology change" for quantum gravity is clear: we should only include those topology changing processes that are dynamically allowed in our theory. Thus, at a rough level, we define the cobordism groups
\begin{equation}\label{Cobordism of Quantum Gravity} \Omega_k^\text{QG} = \left\{ \begin{matrix} \text{Compact, closed} \\ \text{$k$-dimensional backgrounds} \end{matrix} \right\} \Big/ \sim, \end{equation}
where we say two backgrounds $M, N$ are cobordant if we can transition from one to the other by a sequence of dynamically allowed processes in quantum gravity. From the perspective of the $D$-dimensional theory, this process serves as a domain wall between the two different theories, given by compactification on $M$ and $N$.

\vspace{6pt}
\begin{figure}[t]
\centering
\includegraphics[height = 3 in, width = 5 in]{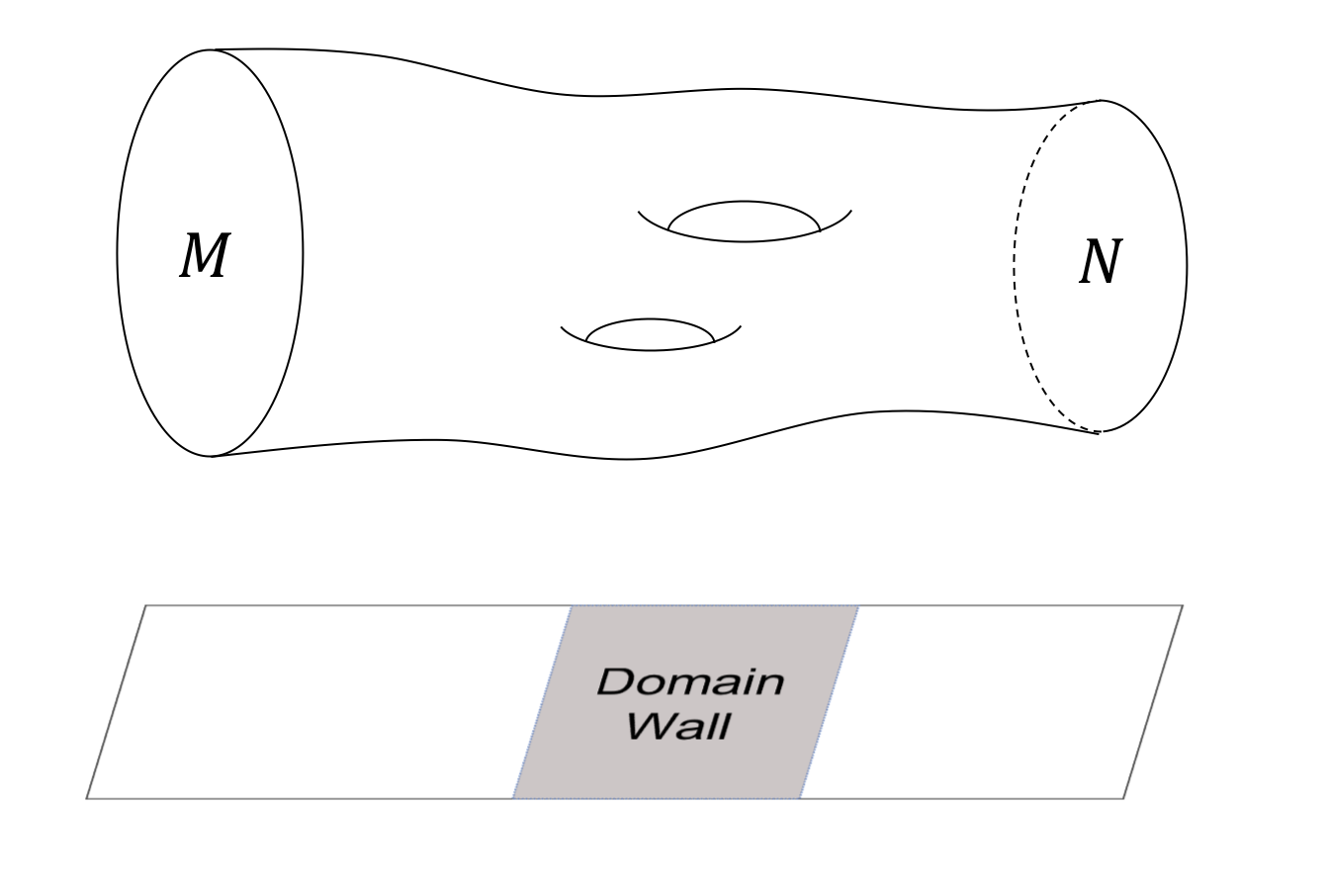}
\caption*{A cobordism of compactification spaces serves as a domain wall.}
\end{figure}

So far, we have only defined $\Omega_k^\text{QG}$ as a set. In order to define a group structure, we consider the disjoint union of two $k$-dimensional backgrounds, $M_1 \sqcup M_2$, and define
\[ [M_1] + [M_2] = [M_1 \sqcup M_2]. \]
That this additive structure is abelian follows from the fact that there is no notion of the ordering of a disjoint union. What we mean by ``compactification on the disjoint union" is really just two separate theories of quantum gravity in $D$-dimensions, namely the theories given by compactification on $M_1$ and $M_2$ respectively. In particular, in our definition (\ref{Cobordism of Quantum Gravity}), we do not require that the $k$-dimensional backgrounds be connected. However, suppose that we are in the situation that the disjoint union $M_1 \sqcup M_2$ can be connected to some other background $N$ by a dynamical process, i.e., suppose we have $M_1 \sqcup M_2 \sim N$. Then we have
\[ [M_1] + [M_2] = [M_1 \sqcup M_2] = [N], \]
and so the group structure encodes the possibility for two separate backgrounds to collide and join. From the perspective of the $D$-dimensional theory, this would behave as a domain wall between the pair of theories defined by compactification on $M_1$ and $M_2$, and the theory defined by compactification on $N$.

\vspace{6pt}
\begin{figure}[!ht]
\centering
\includegraphics[height = 3 in, width = 5 in]{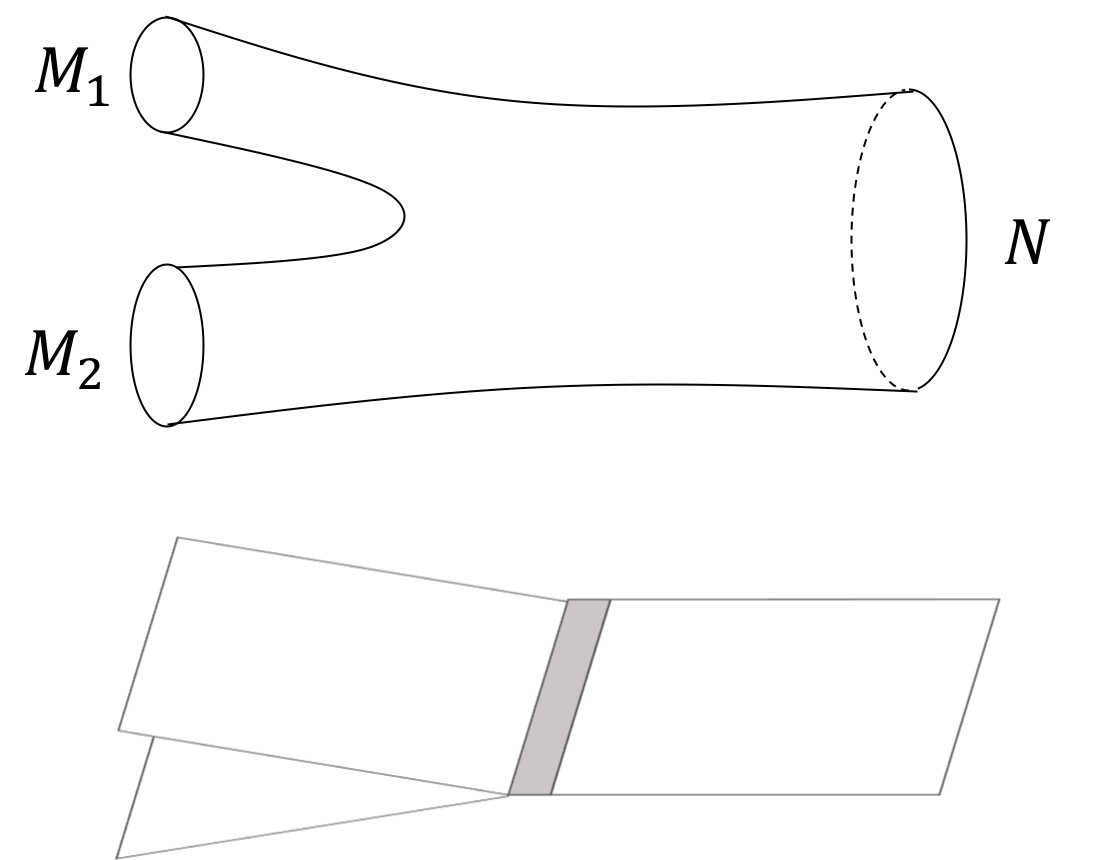}
\vspace{12pt}
\caption*{A cobordism from a disjoint union.}
\end{figure}

The remaining pieces of data needed to define a group structure are a zero element and inverses. Formally, we may define the zero element to be compactification on the ``empty background," which simply means that we have the absence of any $(d - k)$-dimensional theory. If for some background $M$ we have
\[ [M] = 0, \]
then this means that there is a dynamical process by which $M$ may disappear. From the perspective of the $D$-dimensional theory, this process would behave as a domain wall at the edge of spacetime, with nothing beyond. Finally, for inverses, we note that by holding $M$ fixed in time, we obtain a (trivial) cobordism from $M$ to itself. By folding the time in half, we may alternatively view this as a process by which the disjoint union of $M$ and an appropriate orientation-reversal $\overline M$ join and disappear into nothing, and so we have
\[ [M] + [\overline M] = 0. \]
Thus, the set $\Omega_k^\text{QG}$ defined in (\ref{Cobordism of Quantum Gravity}) is indeed an abelian group.

\vspace{6pt}
\begin{figure}[!ht]
\centering
\hspace{24pt}\includegraphics[height = 3 in, width = 5 in]{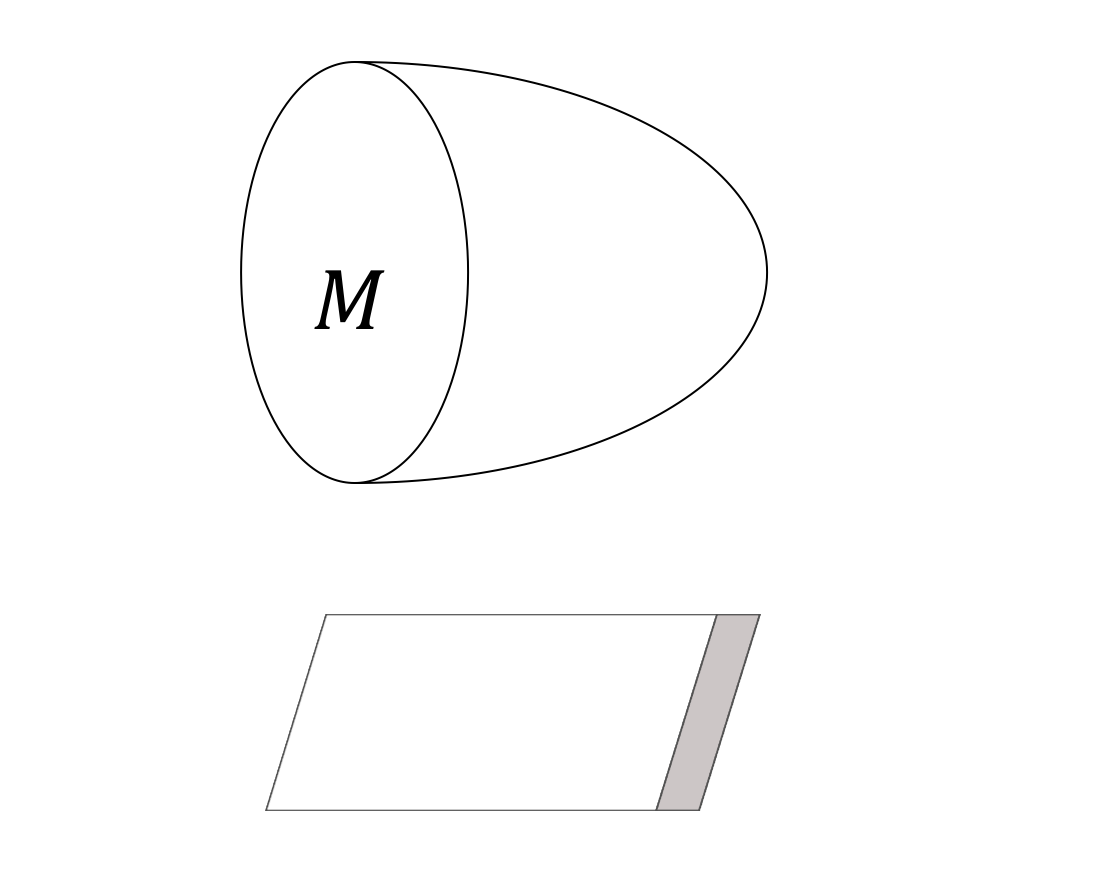}
\caption*{A cobordism to the empty manifold.}
\end{figure}

One alternative perspective on $\Omega_k^\text{QG}$ follows from considering the spacetimes $W$ traced out by the dynamical processes connecting, say, $M$ and $N$ as noncompact $(k+1)$-dimensional backgrounds in their own rights. If we choose a fibering $W \to \RR_s$ for some choice of Euclidean time $s$, then we should have that for $s \to -\oo$ we have that $W \to M \times \RR$, while for $s \to +\oo$ we have $W \to N \times \RR$. In this situation, we have that
\[ \del W = M \sqcup \overline N, \]
which matches more closely onto the usual mathematical notion of cobordism. We note that while $M, N$, and the internal $k$-dimensions of $W$ are allowed to be non-geometric, we are requiring the Euclidean time $s$ to still be geometric in a sense. In this language, we would say that the cobordism class of a background $M$ vanishes,
\[ [M] = 0, \]
if there exists some noncompact $(k + 1)$ dimensional background $W$, such that $M = \del W$. Thus, the condition that $\Omega_k^\text{QG} = 0$ can equivalently be stated as the condition that for any $k$-dimensional compact background of quantum gravity, there must exist a $(k + 1)$ dimensional noncompact background for which it is the boundary. From the perspective of the D-dimensional macroscopic space, this would mean that we expect every quantum gravitational theory to have a $D-1$-dimensional domain wall that it can end on.

\section{Cobordism Invariants in Quantum Gravity}\label{Cobordism Invariants in Quantum Gravity}

In this section we review why global symmetries should be absent in a quantum theory of gravity, and why this implies the vanishing of cobordism classes defined in the previous section. We also clarify the notion of apparent cobordism groups and their predictive power for the existence of topologically non-trivial defects.

\subsection{Conserved Charges and Black Holes}\label{Conserved Charges and Black Holes}

In this section, we review the well-known argument that the process of black hole formation and evaporation should violate the conservation of any supposed global charge in a theory of quantum gravity. The basic argument goes as follows. Suppose that we had a particle carrying the conserved charge, and a large uncharged black hole. We can imagine throwing the particle into the black hole, and waiting for it to evaporate. However, since the early Hawking radiation depends only on the geometry of the black hole event horizon, the Hawking radiation cannot be sensitive to the global charge, and so the charge must remain in the black hole until the end of evaporation. Thus, there are two possibilities. The first is that the conservation of charge is violated, in which case we are done. The second is that the charge remains behind at the end of evaporation in a Planckian remnant. However, since the initial charge was arbitrary, we may thus form an arbitrarily large number of distinct remnants, which signals a sickness of the theory \cite{Susskind:1995da}.

Though this argument is not airtight (it only works for continuous symmetries in particular), it provides good intuition as to why quantum gravity should be free of global symmetries. In particular, it is easy to see where the above argument fails for gauged symmetries: since a charge for a gauge symmetry must be attached to field lines, when it falls into a black hole the field lines remain to pierce the horizon, and so the Hawking radiation can indeed be sensitive to the charge contained within. Put differently, for a gauged symmetry, you can measure the charge inside the black hole from outside the event horizon by using Gauss's law.

Now, we consider what the above argument looks like if the particle we threw into the black hole were a gravitational soliton, in a similar sense to those discussed in \cite{Witten:1985xe}. What we mean by this is the following. Suppose $M^{k}$ is a consistent compactification for our theory. We can imagine forming the connected sum $\RR^{k} \# M$, defined by cutting out a small ball in $\RR^{k}$ and in $M$, and gluing the spherical boundaries together in a standard and smooth way.\footnote{To do this precisely, we would need to specify the metric and all the dynamical fields throughout this gluing process. Since we are only considering global/topological aspects, we will suppress this technicality.} This manifold looks like flat, empty space away from a small region, and so $M$ may serve as a $(d-k-1)$-dimensional defect in our theory. Thus, we get a map
\[ \left\{ \begin{matrix} \text{Compact, closed} \\ \text{$k$-manifolds} \end{matrix} \right\} \rightsquigarrow \left\{ \begin{matrix} (d-k-1)\text{-defects} \end{matrix} \right\}. \]

\vspace{6pt}
\begin{figure}[t]
\centering
\includegraphics[height = 1.5 in]{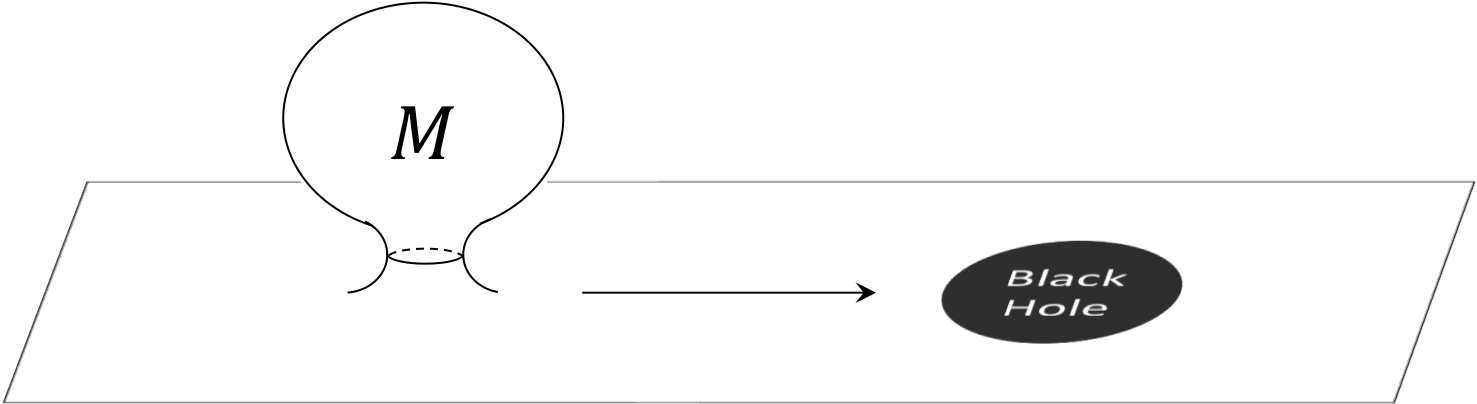}
\vspace{12pt}
\caption*{A gravitational soliton formed from $M$ falls into a black hole.}
\vspace{6pt}
\end{figure}

Now, we may ask whether there are any conserved $(d-k-1)$-form charges associated to the gravitational solitons formed this way. This would mean an invariant of $M^{k}$ that doesn't change under time evolution by a $(k+1)$-dimensional spacetime. This is exactly what we mean by a nontrivial cobordism invariant in $\Omega_{k}^\text{QG}$. Thus, classes in $\Omega_{k}^\text{QG}$ label charges carried by the corresponding defects. The question, then, is whether these charges are global or gauged.

The key for us is that the connected sum $\RR^{k} \# M$ is indistinguishable from empty space away from a localized region. Thus, running the above argument and throwing these defects into a black brane (or a black hole obtained by wrapping them on a toroidal geometry of dimension $(d-k-1)$), we would expect for their charges to be violated by black brane (black hole) evaporation, since we cannot determine the total charge from outside the horizon. Thus, we conclude that these charges are global, not gauged, and thus their presence should be inconsistent with quantum gravity just as any other global charges are inconsistent.\footnote{ We note in passing that the symmetry group in this case would be
\[ \left( \Omega_k^\text{QG} \right)^\vee = \text{Hom} \left( \Omega_k^\text{QG}, U(1) \right), \]
but we do not need to make any reference to the symmetry group, since the notion of conserved charge is more closely related to the physics of black hole evaporation in any case.}

\subsection{Apparent Cobordism Groups}\label{Apparent Cobordism Groups}

While in principle, the statement that
\[ \Omega_k^\text{QG} = 0, \]
could be checked itself in quantum gravity, in practice we are faced with the fact that we don't have a full, non-perturbative description of string theory that makes it clear what are and aren't allowed backgrounds and dynamical processes, especially when it comes to the types of global issues captured by cobordism. Thus, the approach we will take in this paper is one of approximation. Rather than attempt to compute the full-fledged groups $\Omega_k^\text{QG}$ for known string theories, we instead restrict ourselves to much simpler cobordism groups $\Omega_k^{\widetilde{\text{QG}}}$ that involve turning off many fields and ignoring the possibility of singular or non-geometric backgrounds. We then indeed find nontrivial cobordism groups, and we need to know what to do with them.

What a non-vanishing cobordism group of an approximation to a consistent theory is telling us is that the approximation is inconsistent, and that we need to include new ingredients if we want to understand the full theory. Since the presence of a non-vanishing cobordism group implies the existence of a global symmetry, there are two possibilities for the full theory.
\begin{itemize}
\item The first possibility is that this symmetry is actually broken in the full theory. In terms of cobordism, this means that there exist defects that break the symmetry, which we may use to produce new cobordisms between classes that were not previously connected. By including some of these defects, we may refine our approximation. The cobordism for the refined approximation receives a map from the original cobordism,
\[ \Omega_k^{\widetilde{\text{QG}}} \to \Omega_k^{\widetilde{\text{QG}} + \text{defects}}, \]
by mapping onto those manifolds which do not include defects. We say that classes in the kernel of this map are \emph{killed} in the full theory.
\item The second possibility is that this symmetry is actually gauged in the full theory. In terms of cobordism, this means that if we have a $k$-manifold $M$, then the full theory is actually inconsistent when placed on $M$ unless
\[ [M] = 0 \in \Omega_k^{\widetilde{\text{QG}}}, \]
which is to say that the total charge on a compact manifold must vanish. In this case, we may refine our approximation by including the some of the gauge fields for this symmetry. The cobordism for the refined approximation maps into the original cobordism
\[ \Omega_k^{\widetilde{\text{QG}} + \text{gauge fields}} \to \Omega_k^{\widetilde{\text{QG}}}, \]
by forgetting the gauge field. We say that classes in the cokernel of this map are \emph{co-killed} in the full theory.
\end{itemize}
In both cases, we may refine our approximation, and hopefully come closer to the exact theory. However, while adding additional ingredients can remove apparent cobordism classes, it can also produce new ones, and indeed, we only expect to see vanishing cobordism groups at the end of the day when we are able to consider a full, exact theory of quantum gravity.

\vspace{6pt}
\begin{figure}[!ht]
\centering
\includegraphics[height = 2 in]{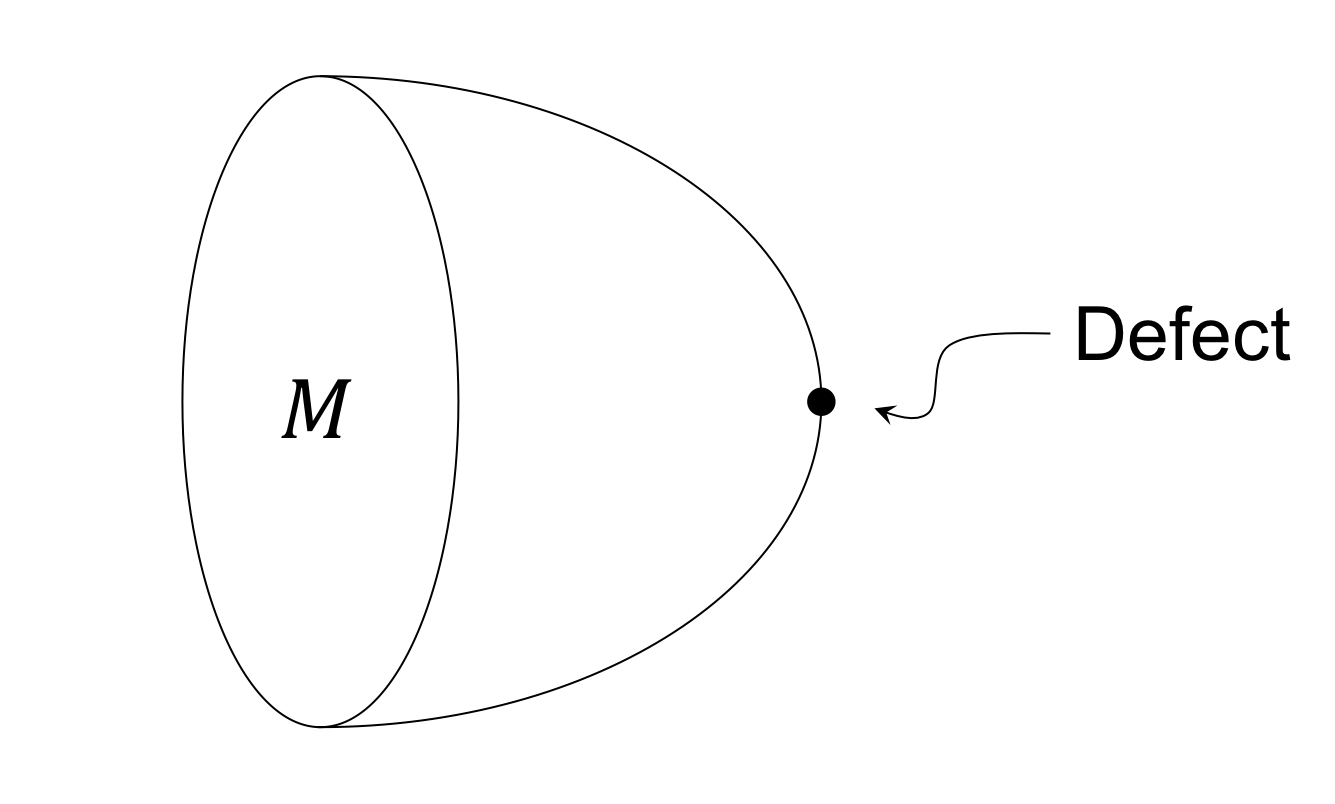}
\caption*{New defects may kill otherwise nontrivial cobordism classes.}
\end{figure}

\section{Examples}\label{Examples}

In this section, we apply the condition
\[ \Omega_k^\text{QG} = 0, \]
to string theory, following the approach described in Section \ref{Apparent Cobordism Groups}. First, we explain how completeness of the spectrum for $p$-form gauge fields is a special case of the condition $\Omega_{p + 1}^\text{QG} = 0$, and thus how the existence of D-branes can be explained by cobordism. Next, we turn to the symmetry group of spacetime itself, and examine the cobordism of spin manifolds as a first approximation to the cobordism of string theory. Finally, we discuss a number of small refinements, in the form of M-theory on nonorienttable manifolds, nontrivial F-theory compactifications of Type IIB, and the appearance of string structures in heterotic string theory.  Moreover, we can interpret the examples found in \cite{Tachikawa:2018njr} as the statement that Type II orientifolds kill the generators of the appropriate cobordism groups that take into account the symmetries $(-1)^{F_L}$ and the orientifold $\Omega$.

\subsection{$p$-Form Gauge Fields and D-Branes}\label{p-Form Gauge Fields and D-Branes}

In this section, we will argue that the condition of completeness of the spectrum for a $p$-form $U(1)$ gauge field is a special case of the vanishing of $\Omega_k^\text{QG}$, and that the required defects are simply the magnetic monopole branes. Of course, it is already well-known that completeness of the spectrum for abelian gauge symmetries follows from absence of global symmetries \cite{Harlow:2018tng, Banks:2010zn}, and so this section is merely a rewriting of known facts in the language of cobordism.

In order to argue that vanishing of cobordism implies the existence of D-branes, we consider the cobordism of oriented\footnote{We restrict to oriented manifolds in order to define the charge (\ref{Charge}). We could extend our discussion to theories defined on nonorientable manifolds provided that $A_p$ is a $p$-form twisted by the orientation bundle of our manifold, such as the M-theory $C$-field \cite{Witten:2016cio}.} manifolds that carry a $p$-form $U(1)$ gauge field $A_p$, with field strength $F_{p+1} = d A_p$. This cobordism group is given by
\[ \Omega_k^{SO,U(1)_p} = \left\{ \begin{matrix} \text{Oriented $k$-manifolds $M^k$} \\ \text{with a p-form gauge field $A_p$} \end{matrix} \right\} \Big/ \text{Cobordism}, \]
the cobordism groups of oriented manifolds with a $p$-form $U(1)$ gauge field $A$. There is a cobordism invariant
\begin{equation}\label{Charge}
Q : \Omega_{p+1}^{SO,U(1)_p} \to \ZZ, \quad Q(M^{p+1}, A_p) = \int_M F_{p+1},
\end{equation}
which measures the total magnetic flux of $F_{p+1}$ through $M$. That this is a cobordism invariant follows from the Bianchi identity $d F_{p+1} = 0$.

The logic of Section \ref{Conserved Charges and Black Holes} tells us that there must be a $(d - p - 2)$-form symmetry in theories of quantum gravity with $p$-form gauge fields $A_p$. This symmetry is well known, and is given by shifting the magnetic dual potential $\wt A_{d - p - 2}$ by a closed $(d - p - 2)$-form. The charged defect under this symmetry is just a localized magnetic flux of $F_{p+1}$, which we identify as the result of taking the connected sum of $\RR^{p+1}$ with a unit of magnetic flux on $S^{p+1}$, as in Section \ref{Conserved Charges and Black Holes}.

What would it mean to break this symmetry, or equivalently, to kill this cobordism class? This would mean that there is some $(d - p - 3)$-defect whose worldvolume is linked by a $(p+1)$-manifold with unit magnetic flux. This is just the definition of a magnetic monopole, and so we have reduced the statement that the cobordism invariant $Q$ should be killed to the statement that there must exist magnetic monopoles for $F_{p+1}$. In particular, we now have that the condition $\Omega_k^\text{QG} = 0$ plus the existence of R-R gauge fields implies the existence of D-branes in Type II string theory. 

\subsection{Spin Cobordism}\label{Spin Cobordism}

As a first approximation to the full cobordism groups of string theory, we will imagine turning off all gauge fields (the most basic properties of which are covered in Section \ref{p-Form Gauge Fields and D-Branes}), and use only the fact that string theory includes fermions. Thus, we consider manifolds with a choice of spin structure, i.e., oriented manifolds with a chosen trivialization of the second Steifel-Whitney classe $w_2 = 0$. The cobordism of spin manifolds is described by the spin cobordism groups $\Omega_k^\Spin$, which are given in the range $0 \leq k < 8$ below \cite{ABP}. The last row of the table lists the generators, where $\pt^+$ is the positively-oriented point, $S^1_p$ is the circle with the periodic spin structures, and $K3$ is the Kummer surface. We only stop at $k = 7$ for convenience, and indeed there are more nonzero cobordism groups for $k \geq 8$, which should have implications for string theory as well, and which are listed in Appendix \ref{Tables}.

\vspace{6pt}
 \begin{figure}[h]
\[ \arraycolsep=5pt\def\arraystretch{1.5} \begin{array}{c|cccccccc}
k & 0 & 1 & 2 & 3 & 4 & 5 & 6 & 7 \\
\hline
\Omega_k^\Spin & \ZZ & \ZZ_2 & \ZZ_2 & 0 & \ZZ & 0 & 0 & 0 \\
\text{Gens} & \pt^+ & S^1_p & S^1_p \times S^1_p & - & K3 & - & - & - \\
 \end{array} \]
 \end{figure}

 \subsubsection{String Theories on $K3$}\label{String Theories on K3}

As an illustrative example, we will start by considering the implications of the group
\[ \Omega_4^\Spin = \ZZ,\]
generated by $K3$. By the reasoning in Section \ref{Conserved Charges and Black Holes}, this means that any theory of quantum gravity with fermions has a potential $(d - 5)$-form symmetry, with charges labeled by $\ZZ$, under which $K3$ has unit charge. In order to describe this symmetry, we will use the fact that a complete cobordism invariant of oriented 4-manifolds $M$ is given by the signature $\sigma(M)$, defined as the signature of the intersection form on the middle cohomology. Rokhlin's theorem states that for a spin 4-manifold $M$, we have
\[ \sigma(M) \equiv 0 \mod 16, \]
and indeed we have
\[ \sigma(K3) = - 16. \]
By the Hirzebruch signature theorem, we have
\[ \sigma(M) = \frac{1}{3} \int_M p_1(R), \]
where
\[ p_1(R) = \frac{1}{8 \pi^2} \tr \left( R \wedge R \right). \]
Suppose we now define the current
\[ J_{p_1} = \frac{1}{48} \star p_1(R) = \frac{1}{384 \pi^2} \star \tr \left( R \wedge R \right). \]
This is a $(d - 4)$-form, and satisfies $d \star J_{p_1} = 0$. Thus, $J_{p_1}$ generates a $(d - 5)$-form $U(1)$ symmetry, which is normalized so that $K3$ has (negative) unit charge, since we have
\[ \int_{K3} \star J_{p_1} = \frac{1}{48} \int_{K3} p_1(K3) = \frac{1}{16} \sigma(K3) = -1. \]

Since we have now identified a potential global symmetry in quantum gravity theories with fermions, we may now ask what happens to this symmetry in each of the different string theories.
\begin{itemize}

\item {\bf M-theory (and Type IIA):} We claim that part of this $U(1)$ symmetry is broken even by smooth manifolds in M-theory (and by extension in Type IIA), because while we have written down the cobordism of oriented manifolds with spin structure, we know \cite{Witten:2016cio} that M-theory makes sense on nonorientable manifolds with $w_2 = 0$, otherwise known as manifolds with pin$^+$ structure. Since we have not allowed nonorientable manifolds in our description of spin cobordism, they may serve as defects that kill off classes which are not the boundary of any orientable manifold with spin structure. In particular, since a nonorientible 4-manifold $M$ only has a $\ZZ_2$-fundamental class $[M] \in H^4(M; \ZZ_2) = \ZZ_2$, we may only define the integral
\[ \int_M \star J_{p_1} \in \ZZ_2, \]
modulo two. Thus, we might expect that spin 4-manifolds $M$ with
\[ [M] \in 2 \ZZ \subset \ZZ = \Omega_4^\Spin, \]
are killed off in pin$^+$ cobordism. Indeed, we have that the map
\[ \left( \Omega_4^\Spin = \ZZ \right) \to \left( \Omega_4^{\Pin^+} = \ZZ_{16} \right), \]
is given by multiplication by eight, and so has kernel $2 \ZZ \subset \ZZ = \Omega_4^\Spin$. This means that smooth pin$^+$ serve as defects to explicitly break $U(1)$ to $\ZZ_2$, which is then extended to a larger $\ZZ_{16}$ symmetry. Put differently, pin$^+$ manifolds produce instantons that only preserve conservation of $U(1)$ charge modulo two, but also allow for charge fractionalization to $1/8$ the previous unit of charge. Of course, there is now the question as to what happens to this new $\ZZ_{16}$ global symmetry, which we address in Section \ref{M-Theory on Nonorientable Manifolds}; the upshot is that this symmetry is exactly killed by known defects in M-theory.

\item {\bf Type IIB:} We defer a detailed description of the class of $K3$ in Type IIB to Section \ref{F-Theory Compactifications of Type IIB}, but we can say that since $K3$ is a perfectly valid background for type IIB string theory, this symmetry must be broken, not gauged. What we will find is that, while conservation of the signature is broken by nontrivial F-theory compactifications, a combination of the signature and the topology of the elliptic fibration seems to be conserved by known processes, and so this example leads to one of our main predictions for a new, non-supersymmetric defect in string theory.

\item {\bf Heterotic:} The $U(1)$ symmetry coming from $\Omega_4^\Spin = \ZZ$ is gauged in heterotic string theory. What this means is that there is a dynamical 6-form $A_6$ in heterotic string theory, together with a coupling of the form
\[ S \supset \int A_6 \wedge \star J_{p_1} = \frac{1}{384 \pi^2} \int A_6 \wedge \tr \left( R \wedge R \right). \]
If we take $A_6 = 24 \wt B$, a multiple of the magnetic dual of the $B$-field, then we do indeed have the term
\begin{equation}
\label{String Structure}
S \supset 24 \int \wt B \wedge \star J_{p_1} = \frac{1}{16 \pi^2} \int \wt B \wedge \tr \left( R \wedge R \right) = \frac{1}{2} \int \wt B \wedge p_1(R),
\end{equation}
in the heterotic action. This is to say that $K3$ carries $24$ units of fivebrane charge, and thus it would be inconsistent to compactify heterotic string theory on $K3$, as is well-known. Of course, there is also a well-known workaround, which is to cancel the fivebrane charge by adding either fivebranes or gauge field instantons, but this takes us away from cobordism without singularities and with gauge fields turned off, so we will not discuss this new class of $K3$ with gauge field flux in this section, though we will come back to it in Section \ref{New Predictions}. One final note (discussed in more detail in Section \ref{Heterotic String and TMF Invariants}) is that the coupling in (\ref{String Structure}) actually implies that with gauge fields turned off, heterotic string theory can only be defined on manifolds with a trivialization of the class $\lambda = p_1/2$, which is also known as a string structure. We have a map
\[ \left( \Omega_4^\String = 0 \right) \to \left( \Omega_4^\Spin = \ZZ \right), \]
and since $\Omega_4^\String$ vanishes, we indeed see that all the nontrivial classes in $\Omega_4^\Spin$ are co-killed by including the gauge field $\wt B$.

\end{itemize}

\subsubsection{String Theories on $S^1_p$}

We now turn to the implications of the group
\[ \Omega_1^\Spin = \ZZ_2, \]
generated by the non-bounding spin structure on $S^1$, namely the periodic spin structure $S^1_p$. This cobordism group signals the potential for a $(d - 2)$-form symmetry, with charge labeled by $\ZZ_2$.
Since $S^1_p$ is a consistent background for all known string theories, we expect this symmetry to always be broken, i.e., we expect there to be backgrounds for all known string theories which have $S^1_p$ as their boundary. We now investigate to what extent this is true within known examples.

\begin{itemize}

\item {\bf M-theory and Type IIA:} As for the class of $K3$, we will see in Section \ref{M-Theory on Nonorientable Manifolds} that the class of $S^1_p$ is also killed by considering nonorientable manifolds, and in particular we have that $S^1_p$ is the boundary of the M\"{o}bius strip as a pin$^+$ manifold, which we may view as a defect in the orientation structure. However, there is another way to kill this class that we will discuss here, because we would like to be able to apply $T$-duality to kill the class of $S^1_p$ in Type IIB. This alternative way is an orientifold background (for Type IIA), and is given by
\begin{equation}
\label{half-cylinder}
\frac{\RR}{\ZZ_2}\times S^1_p, \quad (X, Y) \mapsto \Omega \cdot (-X, Y),
\end{equation}
where $\Omega$ is worldsheet parity. In other words, an O8-plane wrapped on the circle $S^1_p$ at $X = 0$ is the $7$-brane that kills the class of $S^1_p$ in Type IIA. Note that the ``boundary" compactification we mean is at $X \to \oo$, not at $X = 0$, since we view the O8-plane as part of the bulk. For M-theory, we may form the same background, except we would call the defect a Ho\v{r}ava-Witten wall wrapped on $S^1_p$.

\item {\bf Type IIB:} We may apply T-duality to the IIA background (\ref{half-cylinder}) to obtain a IIB background that kills the class of $S^1_p$ for Type IIB, and which is given by the orientifold background
\[ \frac{\RR \times S^1_p}{\ZZ_2}, \quad (X, Y) \mapsto \Omega (-1)^{F_L} \cdot (-X, -Y), \]
where $(-1)^{F_L}$ is the target-space fermion number of the left-moving modes. This background contains two O7-planes at $X = 0, Y = 0, \pi$. Thus, in Type IIB, the class of $S^1_p$ is killed by two O7-planes. We note further that this background is Sen's limit of the compactification of F-theory on $\frac{1}{2} K3$ given by an elliptic fibration over the hemisphere $\frac{1}{2}\CC \PP^1$.  We will see in Section \ref{F-Theory Compactifications of Type IIB} that the trivialization of this cobordism class is a result of passing from spin cobordism to the spin$^c$ cobordism allowed by F-theory, and indeed the map
\[ \left( \Omega_1^\Spin = \ZZ_2 \right) \to \left( \Omega_1^{\Spin^c} = 0 \right), \]
shows that the class of $S^1_p$ is killed by generalizing to spin$^c$ cobordism.

\item {\bf Heterotic:} For heterotic string theory on $S^1_p$, we do not know of a mechanism by which this class is killed.

\end{itemize}

On final note from this section is the for M-theory and Type II string theory, we may take the defects that kill $S^1_p$, wrap them on an additional $S^1_p$, and obtain defects that kill $S^1_p \times S^1_p$. Thus, for these theories, we will not discuss the class of $S^1_p \times S^1_p$ independently. For heterotic string theory, we do not know how to kill the class of $S^1_p \times S^1_p$.

\subsubsection{String Theories on $\pt^+$}

The final nonzero cobordism group $\Omega_k^\Spin$ for $k < 8$ is
\[ \Omega_0^\Spin = \ZZ, \]
which is generated by the class of $\pt^+$, the positively oriented point. Speaking of ``compactifications on $\pt^+$" is just a way of describing the decompactified theory, since we have $M = M \times \pt^+$ for any manifold $M$. Nevertheless, the presence of this cobordism group implies a potential $(d - 1)$-form symmetry of quantum gravity in $d$ spacetime dimensions, with charges labeled by $\ZZ$. The current that generates this symmetry is simply
\[ J_1 = \star 1 = \vol, \]
the volume form. While it may seem silly to discuss this as a symmetry, it fits neatly into the pattern discussed above, and in fact this is the class most closely related to the question of whether, say, Type IIA and IIB admit a finite energy domain wall.

As for the case of $S^1_p$, we know that the known string theories are, by definition, consistent when compactified on $\pt^+$, and so this class should always be killed.

\begin{itemize}

\item {\bf M-theory and Type IIA:} To kill the class of $\pt^+$ in M-theory, we need to identify a domain wall at the end of the world for M-theory. The Ho\v{r}ava-Witten wall is exactly such an object, and shows that this class is killed in M-theory. For Type IIA, we have the O8-plane as a domain wall at the end of the world, which is just the Ho\v{r}ava-Witten wall wrapped on the M-theory circle. We might worry that killing this class automatically kills all the other classes in M-theory, since if we have $W$ with $\del W = \pt^+$, then for any closed $M^k$, we have that
\[ \del \left( W \times M \right) = \del W \times M = \pt^+ \times M = M, \]
and so $M^k$ would already by a boudnary. If this logic were accurate, it would render our statement that $\Omega_k^\text{QG} = 0$ for $k > 0$ in M-theory much less powerful, since it would follow from $\Omega_0^\text{QG} = 0$. However, this is not the case. In particular, as we will see in section \ref{M-Theory on Nonorientable Manifolds}, M-theory can be defined on more general pin$^+$ manifolds, which are defined by the vainishing of $w_2$. It is not true that the product of two pin$^+$ manifolds is pin$^+$, since we have
\[ w_2 \left( M \times N \right) = w_2(M) + w_1(M) w_1(N) + w_2(N), \]
and so if both $M$ and $N$ are pin$^+$ but nonorientable, then $M \times N$ will not be pin$^+$. In terms of the Ho\v{r}ava-Witten wall, this manifests in the fact that the modes living on the wall form a chiral theory in ten dimensions, and so it is inconsistent to wrap the Ho\v{r}ava-Witten wall on a nonorientable manifold.

\item {\bf Type IIB:} While we expect that the class of $\pt^+$ in Type IIB is killed by a domain wall at the end of the world, we do not know any such object in string theory. One objection to such a domain wall might be that Type IIB is chiral, and thus shouldn't admit a boundary wall. This reasoning is flawed. Of course, if a theory has a chiral anomaly, then it cannot admit a boundary wall, but there is no a-priori reason that a non-anomalous, chiral theory cannot have a boundary wall. In particular, consider the string worldsheet as a theory of $(1+1)$-dimensional gravity in the presence of a $B$-field. This theory still has D-branes, which from the perspective of the worldsheet are boundary walls. But because of the nonzero $B$-field, the worldsheet theory is chiral, and thus provides an example of a chiral theory of quantum gravity with a boundary.  Boundaries for more dramatically chiral 2-dimensional theories have also been constructed in the context of asymmetric orbifolds \cite{lust}.

We should also note that one potential way to construct a boundary wall for Type IIB would be to use a domain wall between IIB and IIA, since we could place this domain wall parallel to a IIA $O8$-plane and obtain a domain wall from IIB to nothing. The existence of a domain wall between IIB and IIA has been suggested in \cite{Distler:2009ri}, where the authors considered the mathematically precise definition of Type II orientifolds in the context of twistings of $K$-theory, and found that there was a natural $\ZZ_2$ parameter that controlled whether a given background was IIB or IIA.

\item {\bf Heterotic:} Just as we do not know how to kill the class of $S^1_p$ in heterotic string theory, we also do not know how to kill the class of $\pt^+$. The same comments about boundaries of chiral theories above apply here as well.

\end{itemize}

\subsection{M-Theory on Nonorientable Manifolds}\label{M-Theory on Nonorientable Manifolds}

In this section, we consider a refinement of our description of the cobordism of M-theory, from spin cobordism to pin$^+$ cobordism. The key observation that makes this possible is that M-theory has parity symmetry, and so may be defined on nonorientable manifolds. In order to define spinors on a nonorientable manifold, we need to make a choice about the algebra between reflections and fermion parity. In particular, there are two possibilities, namely
\[ P^2 = 1,\quad P^2 = (-1)^F,\]
where $P$ denotes parity. These two possibilities are respectively the notions of pin$^+$ and pin$^-$ structures \cite{Freed:2016rqq}, defined by the vanishing of $w_2$ and $w_2 + w_1^2$ of the tangent bundle respectively. For M-theory, the relevant structure is pin$^+$ structure \cite{Tachikawa:2018njr, Witten:2016cio}. The pin$^+$ cobordism groups $\Omega_k^{\Pin^+}$ are given in the range $0 \leq k < 9$ below \cite{ABP, Kirby Taylor}. The last row lists the generators, where $\pt$ is the unoriented point, $KB$ is the Klein Bottle, viewed as a twisted $S^1_p$ bundle over $S^1$ with either spin structure, $\RR \PP^n$ is the real projective space, and $\HH \PP^n$ is the quaternionic projective space. As for spin cobordism, there are additional nontrivial cobordism groups for $k \geq 8$, which may be found in Appendix \ref{Tables}.

\vspace{6pt}
\begin{figure}[h]
\[ \arraycolsep=5pt\def\arraystretch{1.5} \begin{array}{c|ccccccccc}
k & 0 & 1 & 2 & 3 & 4 & 5 & 6 & 7 & 8 \\
\hline
\Omega_k^{\Pin^+} & \ZZ_2 & 0 & \ZZ_2 & \ZZ_2 & \ZZ_{16} & 0 & 0 & 0 & \ZZ_2 \times \ZZ_{32} \\
\text{Gens} & \pt & - & KB & KB \times S^1_p & \RR \PP^4 & - & - & - & \HH \PP^2, \RR \PP^8 \\
 \end{array} \]
 \end{figure}
 
As discussed above, the Ho\v{r}ava-Witten wall kills the class of the point. We may view the Ho\v{r}ava-Witten wall as the first example of an M-theory orientifold plane MO9, namely as M-theory on the quotient $\RR/\ZZ_2$. We have further M-orientifolds, given by the MO5-plane and MO1-plane, which are defined by M-theory on the quotients $\RR^5/\ZZ_2$ and $\RR^9/\ZZ_2$ respectively \cite{Dasgupta:1995zm,Witten:1995em}. These defects exactly kill the classes of $\RR \PP^4$ and $\RR \PP^8$ respectively, and so the only remaining classes are the Klein bottle, the product of the Klein bottle and $S^1_p$, and $\HH \PP^2$.
 
We now discuss the class of $\HH \PP^2$. We claim, so long as we turn off all gauge fields, that $\HH \PP^2$ is not a consistent background for M-theory, and so the $\ZZ_2$ symmetry arising from this class is gauged in M-theory, just as the symmetry arising from the class of $K3$ is gauged in heterotic string theory. In order to see this, we recall that there is a tadpole cancelation condition \cite{Beckers} for M-theory on 8-manifolds with $Sp(1) \cdot Sp(2)$ structure \cite{Fiorenza:2019usl} (such as $\HH \PP^2$), given by
\[ N_{M2} + \frac{1}{2} G_4^2(X) = I_8(X), \]
where
\[ I_8(X) = \int_X \frac{p_2(R) - \left(p_1(R)/2\right)^2}{48} \]
We have that
\[ I_8(\HH \PP^2) = \frac{1}{8}, \]
and so as long as $G_4 = 0$, we cannot form a consistent compactification of M-theory on $\HH \PP^2$.

Finally, we have the classes of the Klein bottle and its product with $S^1_p$. For these classes, we can say at least that they must be killed, since we know they are consistent compactifications of M-theory \cite{Aharony:2007du}. However, we do not know by what mechanism they are killed in the full theory. We will discuss these classes more in Section \ref{New Predictions}. One thing we should note is that while fully geometric backgrounds of Type IIA can see the class of M-theory on $KB \times S^1_p$, since this is just IIA on $KB$, the class of M-theory on $KB$ is realized in IIA as an orbifold of $S^1_p$, where we have turned on a Wilson line for the symmetry $(-1)^{F_L}$ of Type IIA \cite{Aharony:2007du}.

\subsection{F-Theory Compactifications of Type IIB}\label{F-Theory Compactifications of Type IIB}

In this section, we discuss how the more general compactifications of Type IIB string theory arising from F-theory refine our notion of cobordism for IIB. In Section \ref{Spin Cobordism}, we saw that the circle $S^1_p$ is killed in IIB by the background,
\[ \frac{\RR \times S^1_p}{\ZZ_2}, \quad (X, Y) \mapsto \Omega (-1)^{F_L} \cdot (-X, -Y), \]
a particular orbifold limit of the compactification of F-theory on $\frac{1}{2}K3$,
\[ \xymatrix{T^2 \ar[r] & \frac{1}{2}K3 \ar[d] \\ & \frac{1}{2} \CC \PP^1} \]
which gives a compactification of IIB on $\frac{1}{2}\CC \PP^1 = \frac{1}{2}S^2$. Even away from this orbifold point, this is a compactification of IIB that kills $S^1_p$, and indeed we might have expected that F-theory would explain how to kill $S^1_p$, since in spin cobordism we have
\[ S^1_p \times T^2 = \left( S^1_p \right)^3 = \del \left( \frac{1}{2} K3 \right), \]
which is exactly this compactification of F-theory.

However, there is still a puzzle, since we know that $S^1_p$ is not the boundary of the unique spin structure on the hemisphere. The resolution is that allowing nontrivial elliptic fibrations in F-theory forces us to generalize from spin cobordism to spin$^c$ cobordism. A spin$^c$ structure is defined as a choice of a lift of $w_2$ from $\ZZ_2$-cohomology to $\ZZ$-cohomology. We denote this lift as $2c_1$. If we further choose a spin$^c$ connection $A^c$, this notation agrees with the standard formula
\[ \int_{\Sigma^2} \frac{F_{A^c}}{2 \pi} = \frac{1}{2} \int_{\Sigma^2} w_2 \mod 2. \]
With this in mind, we may place a half unit flux of $F_{A^c}$ over each hemisphere, which will produce an additional factor of $(-1)$ for fermions moving around the equator, and so the flux of the spin$^c$ connection kills $S^1_p$ in spin$^c$ cobordism.

In order to see why a spin$^c$ structure is the relevant structure for F-theory compactifications of Type IIB, suppose we have an elliptic fibration
\[ \xymatrix{T^2 \ar[r] & X \ar[d] \\ & B} \]
of $X$ over some base $B$. Now, suppose we have a spinor $\psi_X$ on the total space $X$ of the fibration. Locally in the base and away from singular fibers, we may write
\[ \psi_X = \psi_B \tensor \psi_{T^2} = \psi_B \tensor (dz)^{1/2}, \]
using the fact that spinors on $T^2$ are given by sections of a square root $K_{T^2}^{1/2}$ of the canonical bundle. Thus, on the base and away from singular fibers, we see that spinors in the total space descend to spinors in the base valued in a line bundle $L$ such that
\[ L^2 = H^{1, 0}_{\delbar}(T^2). \]
In order to extend this definition over the singular fibers, we note that while neither spinors on the base nor the line bundle $L$ are well defined, the notion of spinor valued in $L$ still makes sense, since we may identify it with the notion of spinor on the total space. Thus, we see that the base naturally carries a spin$^c$ structure, where the integral lift of $w_2(B)$ is given by $2 c_1(L) = c_1(L^2)$. In fact, we can see more. Since the condition of preserving supersymmetry is that the total space $X$ is Calabi-Yau, i.e., $K_X = 0$, we then have that to preserve supersymmetry, we must have
\[ K_X = 2 c_1(L) + K_B = 0. \]

Now, the spin$^c$ cobordism groups are given below for $0 \leq k < 6$ \cite{ABP} (see Appendix \ref{Tables} for a more complete table). In the last row, we list the generators, where $E$ is the Enriques surface (with unique torsion class $x \in H^2(E; \ZZ)$), $H$ is the hyperplane class for the appropriate complex projective space, and we have given two choices of basis for $\Omega_4^{\Spin^c} = \ZZ \times \ZZ$.\footnote{ The first choice is what naturally appears in the twisted Atiyah-Hirzebruch spectral sequence, while the second consists of supersymmetric backgrounds.} For each generator, we have specified the class $2c_1$ as an element of $H^2(-; \ZZ)$, since different choices of lift can correspond to different cobordism classes.

\vspace{6pt}
\begin{figure}[h]
\[ \arraycolsep=5pt\def\arraystretch{1.5} \begin{array}{c|ccccccc}
k & 0 & 1 & 2 & 3 & 4 & 5\\
\hline
\Omega_k^{\Spin^c} & \ZZ & 0 & \ZZ & 0 & \ZZ \times \ZZ & 0 \\
\text{Gens} & \pt^+ & - & (\CC \PP^1, 2H) & - & \begin{matrix} (E, x), (\CC \PP^2, H) \text{ or} \\ (\CC \PP^1 \times \CC \PP^1, 2 H_1 + 2 H_2), (\CC \PP^2, 3 H) \end{matrix} & - \\
 \end{array} \]
 \end{figure}
 
 We do not know of a mechanism by which the first two classes, $\pt^+$ and $(\CC \PP^1, 2H)$, are killed in Type IIB. We already discussed what it would mean to kill the class of $\pt^+$ in Section \ref{Spin Cobordism}, and we will discuss both classes further in Section \ref{New Predictions}.
 
 \subsubsection{F-theory on Calabi-Yau Threefolds}
 
In this section, we discuss the group
\[ \Omega_4^{\Spin^c} = \ZZ \times \ZZ. \]
There are two natural integral cobordism invariants of spin$^c$ 4-manifolds, given by the integral of $(2 c_1)^2$ and the signature $\sigma$. For the generators $(E, x)$ and $(\CC \PP^2, H)$, these take the values
\[ \int_E x^2 = 0, \quad \sigma(E) = -8, \]
\[ \int_{\CC \PP^2} H^2 = 1, \quad \sigma(\CC \PP^2) = 1. \]
Since these classes form a basis, we see that there is a relation
\[ \int_X (2 c_1)^2 = \sigma(X) \mod 8, \]
but that the characteristic numbers are otherwise unconstrained, and form complete cobordism invariants. Thus, we may express the supersymmetric generators in terms of $(E, x)$ and $(\CC \PP^2, H)$ just by calculating their invariants. We have
\[ \int_{\CC \PP^1 \times \CC \PP^1} \left( 2 H_1 + 2 H_2 \right)^2 = 8, \quad \sigma(\CC \PP^1 \times \CC \PP^1) = 0, \]
\[ \int_{\CC \PP^2} (3 H)^2 = 9, \quad \sigma(\CC \PP^2) = 1. \]
Thus, we have
\[ \begin{pmatrix} (\CC \PP^1 \times \CC \PP^1, 2 H_1 + 2 H_2) \\ (\CC \PP^2, 3 H) \end{pmatrix} = \begin{pmatrix} 1 & 8 \\ 1 & 9 \end{pmatrix} \begin{pmatrix} (E, x) \\ (\CC \PP^2, H) \end{pmatrix}, \]
and since
\[ \det \begin{pmatrix} 1 & 8 \\ 1 & 9 \end{pmatrix} = 1, \]
this is an invertible transformation between two integral bases. The final piece of structure in $\Omega_k^{\Spin^c}$ we should note is that that class of $K3$ with $2c_1 = 0$ is given by $2 (E, x)$, as can be seen by matching cobordism invariants.

Now, since we have $\Omega_4^{\Spin^c} \neq 0$, we expect a potential 5-form symmetry of Type IIB string theory, with charges labeled by $\ZZ \times \ZZ$.
This extends the potential $U(1)$ symmetry we saw in spin cobordism. What is the fate of this symmetry group in Type IIB string theory? We claim that part of it, at least, is broken by known defects. In order to see this, we recall the example of F-theory on a $K3$-fibered Calabi-Yau threefold $CY_3$,
\[ \xymatrix{K3 \ar[r] & CY_3 \ar[d] \\ & \CC \PP^1} \]
There are several possibilities for the base $B$ of this elliptic fibration, and for particular choices of the moduli, we have $B = \mathbb{F}_n$ for $-12 \leq n \leq 12$, where $\mathbb{F}_n$ is the del Pezzo surface.

However, these compactifications are all connected \cite{Witten:1996qb,Morrison:1996pp} via transitions involving the E-string CFT. An easy way to see this is to fiber the duality between F-theory on $K3$ and heterotic string theory on $T^2$ in order to view this compactification as heterotic string theory on
\[ \xymatrix{T^2 \ar[r] & K3 \ar[d] \\ & \CC \PP^1} \]
If we use $E_8 \times E_8$ heterotic string theory, then the parameter $n$ labeling the different del Pezzo surfaces is identified with the division of the 24 required gauge instantons as $(12 - n, 12 + n)$ between the first and second $E_8$. Of course, these heterotic backgrounds are connected. In the M-theory description, this corresponds to shrinking an instanton to zero size, pulling it off the wall as a fivebrane, and allowing it to move across and dissolve in the other wall. In terms of Type IIB, this corresponds to blowing up $\mathbb{F}_n$ at a point followed by blowing down a curve.

Thus, we have learned that the process of blowing up a point is a dynamical process in Type IIB string theory compactified on a 4-manifold. Since blowing up is equivalent topologically to taking a connected sum with $\overline{\CC \PP^2}$, we see that the process of blowing up kills some class represented by $\overline{\CC \PP^2}$. In fact, because this process preserves supersymmetry, we know that we will have
\[ 2 c_1 = K_{Bl(X)} = K_X - e, \]
where $e$ is the exceptional divisor. Thus, we see that the class which is killed is
\[ (\overline{\CC \PP^2}, -H) = - (\CC \PP^2, H), \]
which leaves only a single $\ZZ$ in cobordism, generated by $(E, x)$. Further, we have for the supersymmetric classes
\[ (E, x) \equiv (\CC \PP^1 \times \CC \PP^1, 2 H_1 + 2 H_2) \equiv (\CC \PP^2, 3 H) \mod (\CC \PP^2, H), \]
in cobordism, and so all the supersymmetric backgrounds are connected in cobordism by the process of blowing up and down.

We do not know how to kill this remaining class by known processes in Type IIB. However, we note that one representative is given by the product of two copies of the generator $(\CC \PP^1, 2H)$ of $\Omega_2^{\Spin^c} = \ZZ$. Thus, if this class is killed by some defect with spatial dimension 6 that may be consistently wrapped on an additional $(\CC \PP^1, 2H)$, then we will have also killed the class of the product. We will discuss this class more in Section \ref{New Predictions}.

\subsection{Heterotic String and String Cobordism}\label{Heterotic String and TMF Invariants}

Finally, in this section, we discuss how the constraint
\[ dH = \frac{p_1(R) - p_1(F)}{2}, \]
in heterotic string theory modifies the cobordism groups in the case that the gauge field $F$ vanishes. In this case, as discussed in Section \ref{String Theories on K3}, this implies that the class $\lambda = p_1/2$ defined on a spin manifold must vanish for allowed backgrounds of heterotic string theory. In mathematics, given a manifold with spin structure, a choice of trivialization of the class $\lambda$ is referred to as a string structure, and so we consider the cobordism groups $\Omega_k^\String$ of string manifolds, which are given in the range $0 \leq k < 8$ below \cite{Giambalvo} (see Appendix \ref{Tables} for a more complete table). We have generators $S^1_p$ and $S^3_H$,where $H$ subscript denotes the fact that there is a unit of H-flux (alternatively, a unit of framing).

\vspace{6pt}
\begin{figure}[h]
\[ \arraycolsep=5pt\def\arraystretch{1.5} \begin{array}{c|cccccccc}
k & 0 & 1 & 2 & 3 & 4 & 5 & 6 & 7 \\
\hline
\Omega_k^\String & \ZZ & \ZZ_2 & \ZZ_2 & \ZZ_{24} & 0 & 0 & \ZZ_2 & 0 \\
\text{Gens} & \pt^+ & S^1_p & S^1_p \times S^1_p & S^3_H & - & - & S^3_H \times S^3_H & - \\
 \end{array} \]
 \end{figure}
 
We do not know how the first three classes of $\pt^+$, $S^1_p$, and $S^1_p \times S^1_p$ are killed. Note that by duality, heterotic string theory on $S^1_p \times S^1_p$ is equivalent to IIB on $(\CC \PP^1, 2 H)$, another class which we didn't know how to kill. However, it is possible to kill the class of $S^3_H$ \cite{Gukov} and indeed this class is killed by the fivebrane of heterotic string theory, which is surrounded by an $S^3$ with one unit of $H$-flux.

\section{New Predictions}\label{New Predictions}

Throughout Section \ref{Examples}, we identified a number of examples of classes represented by consistent string backgrounds that are not killed by any mechanism of which we know in string theory. Thus, the condition $\Omega_k^\text{QG} = 0$ predicts the existence of a number of new objects in string theory that have not been constructed to our knowledge. We record these classes by the remaining number of large dimensions, as well as the dualities between them, in Figure \ref{Remaining Classes}. We make no claim as to completeness, and indeed there are some easily identified classes (such as M-theory on the M\"{o}bius strip) that we do not list since they fall outside of the very restrictive cobordism groups we considered in Section \ref{Examples}, where we have turned off gauge fields and ignored cobordism with singularities. We do not omit them to imply that they are killed by known objects in string theory, but because we have not computed the correct cobordism groups to discuss them.

\vspace{6pt}
\begin{figure}[t]
\begin{center}
\arraycolsep=5pt\def\arraystretch{1.5} \begin{tabular}{|c|}
\hline
{\bf Dimension 10} \\
\hline \hline
Type IIB on $\pt^+$\\
\hline
Heterotic $E_8 \times E_8$ or $\Spin(32)/\ZZ_2$ on $\pt^+$ \\
\hline \hline
{\bf Dimension 9} \\
\hline \hline
Heterotic on $S^1_p$ \\
\hline
M-theory on $KB$ \\
\hline \hline
{\bf Dimension 8} \\
\hline \hline
Heterotic on $S^1_p \times S^1_p \longleftrightarrow$ Type IIB on $(\CC \PP^1, 2H)$ \\
\hline
M-theory on $KB \times S^1_p$ \\
\hline \hline
{\bf Dimension 6} \\
\hline \hline
$\begin{matrix} \text{Type IIB on $(\CC \PP^1 \times \CC \PP^1, 2 H_1 + 2 H_2) \longleftrightarrow$ Heterotic on $K3$} \\ \text{or Type IIB on $(K3, 0)$} \end{matrix}$ \\
\hline
 \end{tabular}
 \end{center}
 \caption{Remaining cobordism classes.}
 \label{Remaining Classes}
 \end{figure}
 
What are we to make of these so-far unkilled classes? As we will argue more precisely below, any defect that kills any one of the classes listed in Figure \ref{Remaining Classes} must necessarily break all supersymmetry. While we have a good understanding of the supersymmetric defects in string theory, we have a much less complete picture of which non-supersymmetric defects exist, and so we view this as a prediction of new non-supersymmetric defects which kill all the classes in Figure \ref{Remaining Classes}. Since such a defect would have an otherwise nontrivial boundary, its presence can be detected from far away, as discussed below, and so at least a topological piece of the predicted defect must be stable against decay. 

\subsection{Violation of Supersymmetry}

In this section, we argue that all the predicted defects that we have not been able to identify (which kill the classes listed in Figure \ref{Remaining Classes}) must necessarily break all supersymmetry. In order to do this, we note that compactification on all the classes in Figure \ref{Remaining Classes} yield quantum gravity theories of two types. The first are theories in dimensions $d = 10$, 9, 8, or 6 with (potentially chiral) $\mathcal{N} = 1$ supersymmetry. The second are theories in dimensions $d = 10$ or 6 with chiral $\mathcal{N} = (2, 0)$ supersymmetry. What we will argue is that a domain wall at the end of the world for any of these theories must break all supersymmetry; since a defect in the higher dimensional theory would yield such a domain wall, it must also break all supersymmetry.

First of all, we note that the presence of a domain wall must break at least part of the $d$-dimensional supersymmetry, since it breaks translation symmetry in the perpendicular direction. However, it does preserve $(d-1)$-dimensional Poincare invariance, and so we may ask that it preserve some amount of supersymmetry in $(d - 1)$-dimensions. For the $\mathcal{N} = 1$ theories in $d = 10$, 9, 8, or 6, we note that dimensional reduction to $(d - 1) = 9$, 8, 7, or 5 yields $\mathcal{N} = 1$ supersymmetry in the lower dimensional theory, and so there is no intermediate $(d - 1)$-dimensional supersymmetry between preserving all the $d$-dimensional supersymmetry and preserving none of it, and so these defects must break all supersymmetry.

For the $d = 10$ or 6 dimensional theories with $\mathcal{N} = (2, 0)$ supersymmetry, the reduction to $(d - 1) = 9$ or 5 dimensions yields $(d - 1)$-dimensional $\mathcal{N} = 2$ supersymmetry, and so there seems to be the possibility of breaking only half the $(d - 1)$-dimensional supersymmetry. However, this is not the case. Suppose the two supercharges in $d$-dimensions are $Q_1$ and $Q_2$. Since in these cases the $(d - 1)$-dimensional Lorentz group acts faithfully on the $d$-dimensional spinor representations, if we are the preserve $\mathcal{N} = 1$ supersymmetry in $(d - 1)$-dimensions, then we must preserve some linear combination
\[ Q = a Q_1 + b Q_2. \]
However, since $Q_1, Q_2$ have the same chirality, $Q$ is thus also a well defined $(1, 0)$ supercharge in $d=10$ or $6$ dimensions, which we argued above may not be preserved by the domain wall. Thus the defects for classes with $\mathcal{N} = (2, 0)$ supersymmetries must also break all supersymmetry. \footnote{Note that this argument doesn't work for $\mathcal{N} = (1, 1)$ supersymmetry in $d = 10$ or 6 dimensions, since say $Q = Q_1 + Q_2$ being composed of two irreducible representations of the spinor group do not lead to translations perpendicular to the wall. Indeed this combination of supercharges is preserved by the Type IIA O8-plane.}

\subsection{Properties of Predicted Defects}

Now that we have argued that the condition $\Omega_k^\text{QG} = 0$ predicts a number of new defects in string theory, the natural next question is what properties the required defects must have. For example, must they couple to gauge fields? Must they carry additional dynamical modes on their worldvolumes? In general, we cannot say much more besides what is immediately implied by the condition that they must be linked by a geometry which is otherwise nontrivial in cobordism.

One important mathematical consequence of this fact is that the predicted defects carry a topological charge, which we claim is in some sense ``gauged". Indeed, suppose we have a defect which kills off the generator of an apparent cobordism group $\Omega_k^{\widetilde{\text{QG}}} = \ZZ_n$, for $2 \leq n \leq \oo$. We can then define a $\ZZ_n$ topological charge, which we claim must vanish on any closed manifold. Indeed, suppose we attempt to place a number $m$ of the new defect at points on a closed $(k + 1)$-dimensional manifold. Cutting out small open neighborhoods of the defects, we obtain a $(k + 1)$-manifold without defects, whose boundary is given by $m$ copies of the generator of $\Omega_k^{\widetilde{\text{QG}}} = \ZZ_n$. This is only possible if $m \equiv 0 \mod n$, and thus the $\ZZ_n$ topological charge must vanish on any closed manifold. One example of this is heterotic string theory on $K3$ with 24 fivebranes, which serve as defects to kill off $\Omega_3^\String = \ZZ_{24}$ as discussed in Section \ref{Heterotic String and TMF Invariants}.

While this topological charge is ``gauged" in the sense that the total charge must vanish on a closed manifold, we are not claiming that the predicted defects come with new dynamical gauge fields. One way to resolve this potential tension is to recall from Section \ref{Conserved Charges and Black Holes} why we expect any conserved charge in quantum gravity to be gauged. The key point was that for gauged charges, we can measure the total charge inside a black hole by measuring the flux of the gauge field through the horizon. While a defect killing off an apparent cobordism group $\Omega_k^{\widetilde{\text{QG}}}$ may not couple to an independent gauge field, if we imagine throwing this defect into a black hole, we can still detect its presence from outside the black hole, since the horizon will now realize a nonzero class in $\Omega_k^{\widetilde{\text{QG}}}$. For example, the defect killing Type IIB on $(\CC \PP^1, 2 H)$ is a non-supersymmetric junction of 24 $(p,q)$ 7-branes, which does not couple to a new gauge field in Type IIB, but whose presence inside a black hole could be inferred from counting the number of 7-branes piercing the horizon.

\vspace{6pt}
\begin{figure}[!ht]
\centering
\includegraphics[height = 2 in]{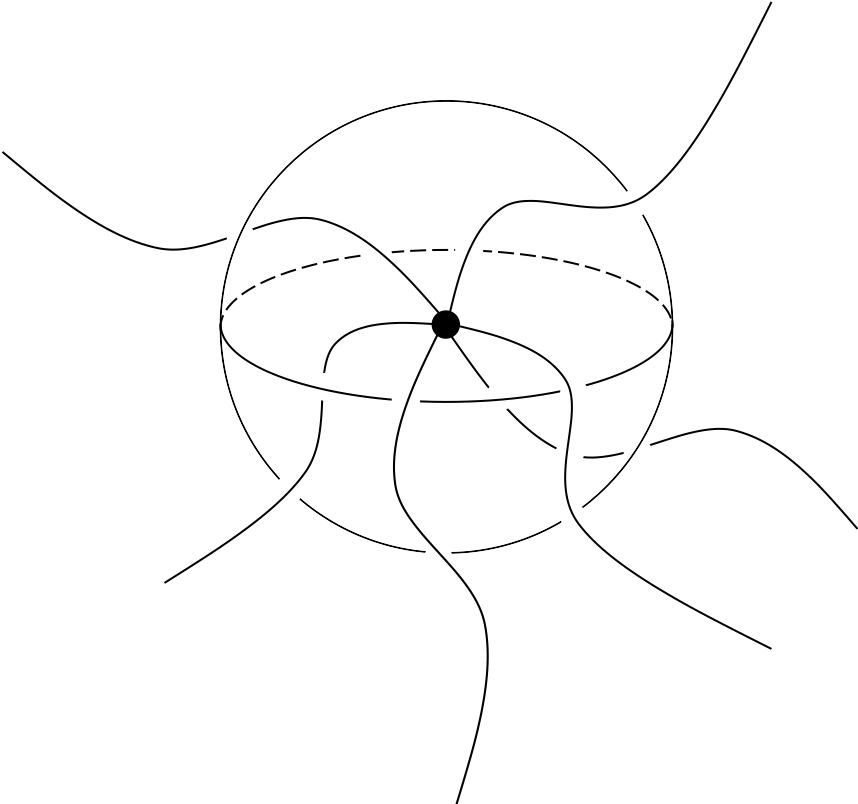}
\caption*{A non-supersymmetric junction of 24 $(p, q)$ 7-branes.}
\vspace{12pt}
\end{figure}

\section{Conclusion}\label{Conclusion}

In this paper, we have argued that the triviality of the cobordism groups of quantum gravity follows from the absence of global symmetries, and thus $\Omega_k^\text{QG} = 0$ is a consistency condition on allowed theories of quantum gravity. Though we have only taken some first steps towards analyzing the cobordism groups of the most general allowed backgrounds in string theory, we have already seen that this condition places highly nontrivial constraints on the allowed objects in string theory, and in particular predicts the existence of a number of new non-supersymmetric defects. It would be very interesting to perform a similar analysis that takes into account more of the features of string theory. One simple extension would be to include the M-theory $C$-field and the induced $\mathfrak{m}^c$ structure \cite{Witten:2016cio}, as the appropriate cobordism groups have been recently computed \cite{Freed:2019sco}. Another natural extension would be to compute the cobordism groups including the new predicted defects. This could potentially result in new cobordism classes, which would then need to be killed off by yet more defects! This cycle could in principle continue until we converge at the correct definition of quantum gravity which would in particular have $\Omega_k^{\text QG}=0$.

\section*{Acknowledgments}
We would like to thank Alek Bedroya, Thomas Dumitrescu, Dan Freed, Sergei Gukov, Arthur Hebecker, Simeon Hellerman, Mike Hopkins, Dieter L\"ust, Sunil Mukhi, David Morrison, Du Pei, Pavel Putrov and Pablo Soler for useful discussions.
We would also like to thank the SCGP where part of this work was done in the 2019 summer workshop.

The research of CV is supported
in part by the NSF grant PHY-1719924 and by a grant
from the Simons Foundation (602883, CV).
This material is based upon work supported by the National Science Foundation Graduate
Research Fellowship Program under Grant No. DGE1745303. Any opinions,
findings, and conclusions or recommendations expressed in this material are those of the
authors and do not necessarily reflect the views of the National Science Foundation.

\begin{appendix}

\section{Table of Cobordism Groups}\label{Tables}

Here, we present a more complete table of the cobordism groups used throughout this paper for reference, up to $k = 10$. The spin and spin$^c$ cobordism groups were initially computed in \cite{ABP}. The pin$^+$ cobordism groups were computed in \cite{Kirby Taylor}, and the string cobordism groups in \cite{Giambalvo}. Good overviews of the calculation of these groups and their roles in physics can be found in \cite{Freed:2016rqq, Garcia-Etxebarria:2018ajm, Wan:2018bns}.

\vspace{6pt}
\begin{figure}[h]
\[ \arraycolsep=5pt\def\arraystretch{1.5} \begin{array}{c|ccccccccccc}
k & 0 & 1 & 2 & 3 & 4 & 5 & 6 & 7 & 8 & 9 & 10 \\
\hline
\Omega_k^\Spin & \ZZ & \ZZ_2 & \ZZ_2 & 0 & \ZZ & 0 & 0 & 0 & \ZZ^2 & \ZZ_2^2 & \ZZ_2^3 \\
\Omega_k^{\Pin^+} & \ZZ_2 & 0 & \ZZ_2 & \ZZ_2 & \ZZ_{16} & 0 & 0 & 0 & \ZZ_2 \times \ZZ_{32} & 0 & \ZZ_2^3 \\
\Omega_k^{\Spin^c} & \ZZ & 0 & \ZZ & 0 & \ZZ^2 & 0 & \ZZ^2 & 0 & \ZZ^4 & 0 & \ZZ^4 \\
\Omega_k^\String & \ZZ & \ZZ_2 & \ZZ_2 & \ZZ_{24} & 0 & 0 & \ZZ^2 & 0 & \ZZ_2 \times \ZZ & \ZZ_2^2 & \ZZ_6 \\
 \end{array} \]
 \end{figure}

\end{appendix}


\begin{thebibliography}{99}

\bibitem{Reid}
M. Reid, ``The moduli space of 3-folds with K = 0 may nevertheless be irreducible,"
Math. Ann. {\bf 278}, 329 (1987). 

 \bibitem{Gaiotto:2014kfa} 
 D.~Gaiotto, A.~Kapustin, N.~Seiberg and B.~Willett,
 ``Generalized Global Symmetries,''
 JHEP {\bf 1502}, 172 (2015)
 [hep-th/1412.5148].

\bibitem{Harlow:2018tng} 
 D.~Harlow and H.~Ooguri,
 ``Symmetries in quantum field theory and quantum gravity,''
 [hep-th/1810.05338].

 \bibitem{Kaidi:2019pzj} 
 J.~Kaidi, J.~Parra-Martinez and Y.~Tachikawa,
 ``GSO projections via SPT phases,''
 [hep-th/1908.04805].

 \bibitem{Susskind:1995da} 
 L.~Susskind,
 ``Trouble for remnants,''
 [hep-th/9501106].

 \bibitem{Witten:1985xe} 
 E.~Witten,
 ``Global Gravitational Anomalies,''
 Commun.\ Math.\ Phys.\ {\bf 100}, 197 (1985).

 \bibitem{Tachikawa:2018njr} 
 Y.~Tachikawa and K.~Yonekura,
 ``Why are fractional charges of orientifolds compatible with Dirac quantization?''
 [hep-th/1805.02772].

\bibitem{Banks:2010zn} 
 T.~Banks and N.~Seiberg,
 ``Symmetries and Strings in Field Theory and Gravity,''
 Phys.\ Rev.\ D {\bf 83}, 084019 (2011)
 [hep-th/1011.5120].
  
 \bibitem{ABP}
 D. W. Anderson, E. H. Brown, Jr. and F. P. Peterson,
 ``The structure of the Spin cobordism ring,"
 Ann.\ of Math.\ {\bf 86}, 271-298 (1967).
 
  \bibitem{lust}
  R.~Blumenhagen, L.~Gorlich, B.~Kors and D.~Lust,
``Asymmetric orbifolds, noncommutative geometry and type I string vacua,''
  Nucl.\ Phys.\ B {\bf 582}, 44 (2000)
  [hep-th/0003024].
 
 \bibitem{Distler:2009ri} 
 J.~Distler, D.~S.~Freed and G.~W.~Moore,
 ``Orientifold Precis,''
 In *Sati, Hisham (ed.) et al.: Mathematical Foundations of Quantum Field theory and Perturbative String Theory* 159--171
 [hep-th/0906.0795].

\bibitem{Freed:2016rqq} 
 D.~S.~Freed and M.~J.~Hopkins,
 ``Reflection positivity and invertible topological phases,''
 [hep-th/1604.06527].
 
  \bibitem{Witten:2016cio} 
 E.~Witten,
 ``The `Parity' Anomaly On An nonorienttable Manifold,''
 Phys.\ Rev.\ B {\bf 94}, no. 19, 195150 (2016)
 [hep-th/1605.02391].
 
 \bibitem{Kirby Taylor}
 R. C. Kirby and L. R. Taylor,
 ``A calculation of Pin$^+$ bordism groups,"
 Comment.\ Math.\ Helvetici\ {\bf 65}, 434--447 (1990).
 
 \bibitem{Dasgupta:1995zm} 
  K.~Dasgupta and S.~Mukhi,
 ``Orbifolds of M theory,''
  Nucl.\ Phys.\ B {\bf 465}, 399 (1996)
  [hep-th/9512196].
  
   \bibitem{Witten:1995em} 
  E.~Witten,
  ``Five-branes and M theory on an orbifold,''
  Nucl.\ Phys.\ B {\bf 463}, 383 (1996)
  [hep-th/9512219].
 
  \bibitem{Beckers} 
 K.~Becker and M.~Becker,
 ``M theory on eight manifolds,''
 Nucl.\ Phys.\ B {\bf 477}, 155 (1996)
[hep-th/9605053].

\bibitem{Fiorenza:2019usl} 
  D.~Fiorenza, H.~Sati and U.~Schreiber,
  ``Twisted Cohomotopy implies M-Theory anomaly cancellation,''
  [hep-th/1904.10207].
 
\bibitem{Aharony:2007du} 
 O.~Aharony, Z.~Komargodski and A.~Patir,
 ``The Moduli space and M(atrix) theory of 9d N = 1 backgrounds of M/string theory,''
 JHEP {\bf 0705}, 073 (2007)
 [hep-th/0702195].

 \bibitem{Witten:1996qb} 
 E.~Witten,
 ``Phase transitions in M theory and F theory,''
 Nucl.\ Phys.\ B {\bf 471}, 195 (1996)
[hep-th/9603150].
 
 \bibitem{Morrison:1996pp} 
 D.~R.~Morrison and C.~Vafa,
 ``Compactifications of F theory on Calabi-Yau threefolds 2,''
 Nucl.\ Phys.\ B {\bf 476}, 437 (1996)
[hep-th/9603161].
 
 \bibitem{Giambalvo}
 V.~Giambalvo,
 ``On $\langle 8 \rangle$-cobordism,"
 Illinois\ J.\ Math. {\bf 15}, no. 4, 533--541 (1971).
 
 \bibitem{Gukov}
  S. Gukov, D. Pei, P. Putrov and C. Vafa, in attempts to represent TMF classes by branes, unpublished.
  
  \bibitem{Freed:2019sco} 
 D.~S.~Freed and M.~J.~Hopkins,
 ``M-Theory anomaly cancellation,''
 [hep-th/1908.09916].
  
  \bibitem{Garcia-Etxebarria:2018ajm} 
 I.~Garcia-Etxebarria and M.~Montero,
 ``Dai-Freed anomalies in particle physics,''
 JHEP {\bf 1908}, 003 (2019)
 [hep-th/1808.00009].
 
 \bibitem{Wan:2018bns} 
  Z.~Wan and J.~Wang,
  ``Higher Anomalies, Higher Symmetries, and Cobordisms I: Classification of Higher-Symmetry-Protected Topological States and Their Boundary Fermionic/Bosonic Anomalies via a Generalized Cobordism Theory,''
  Ann.\ Math.\ Sci.\ Appl.\  {\bf 4}, no. 2, 107 (2019)
  [hep-th/1812.11967].
 



\end{thebibliography}
\end{document}